# Use and Misuse of the Term Experiment in Mining Software Repositories Research

## Claudia Ayala, Burak Turhan, Xavier Franch, and Natalia Juristo

**Abstract**— The significant momentum and importance of Mining Software Repositories (MSR) in Software Engineering (SE) has fostered new opportunities and challenges for extensive empirical research. However, MSR researchers seem to struggle to characterize the empirical methods they use into the existing empirical SE body of knowledge. This is especially the case of MSR experiments. To provide evidence on the special characteristics of MSR experiments and their differences with experiments traditionally acknowledged in SE so far, we elicited the hallmarks that differentiate an experiment from other types of empirical studies and characterized the hallmarks and types of experiments in MSR. We analyzed MSR literature obtained from a small-scale systematic mapping study to assess the use of the term experiment in MSR. We found that 19% of the papers claiming to be an experiment are indeed not an experiment at all but also observational studies, so they use the term in a misleading way. From the remaining 81% of the papers, only one of them refers to a genuine controlled experiment while the others stand for experiments with limited control. MSR researchers tend to overlook such limitations, compromising the interpretation of the results of their studies. We provide recommendations and insights to support the improvement of MSR experiments.

—————————— ◆ ——————————

# 1 INTRODUCTION

EMPIRICAL studies in software engineering (SE) have become popular and have grown in maturity and rigor [14],[67],[73]. To support experimentation in SE, the empirical SE community has devoted efforts to learn, adapt, and mature a great variety of empirical methods and instruments such as experiments, case studies, surveys and systematic literature reviews used in other consolidated experimental disciplines [26], [28], [31], [32], [33], [34], [46], [51], [53], [58], [62], [66], [67]. Along this path, several challenges and lines of research have emerged to face empirical SE endeavors. Nowadays, the significant momentum and importance of mining software repositories (MSR) area is fostering new opportunities and new challenges for extensive empirical research in SE.

MSR researchers analyze the rich data available in software repositories to uncover information about software systems, their development and developers. Software repositories such as version control systems, archived communications between project stakeholders, build logs, test executions, and issue-tracking systems are used to help manage the progress of software projects. These repositories are often related to Open Source Software (OSS) projects but also to internal company repositories [38], [44], [48]. The premise of MSR is that empirical and systematic investigations of repositories will shed new light on a wide spectrum of processes and changes that occur over time by uncovering pertinent information, relationships, or trends [29].

The literature has discussed the challenges that MSR research imposes on SE researchers, as they usually need to work with new types of data, tools or analysis techniques [17], [22], [23], [73]. Several studies have highlighted the issues in analyzing and interpreting MSR research results [12], [25], [30], [39], [44], [45], [73]. However, the challenges imposed from a methodological perspective have not been thoroughly studied yet.

We have observed that MSR researchers seem to struggle to characterize the empirical methods they use into the existing empirical SE body of knowledge. A great deal of MSR publications use the generic term "empirical study" to avoid further debates on the empirical methods used. We think that such confusion mainly comes from the fact that the nature of software repositories' data does not directly match to the types of empirical studies acknowledged so far in SE [14], [67]. This has generated some improper uses of the empirical SE terminology, but more importantly, unclear assumptions on the studies' design and results interpretation. To the best of our knowledge, there is no attempt to study the specific characteristics of MSR research in order to shed light on the proper use of SE empirical methods and terminology in MSR studies. In this paper, we focus on the assessment of a specific type of empirical study: experiments. Our goal is to provide evidence on the special characteristics of experiments in MSR to contribute to a better understanding of key methodological aspects for improving the design and interpretation of MSR experiments, as well as for raising the awareness on the need of recognizing the differences between MSR experiments with the current concept of "experiment" used in SE so far.

To do so, we performed an in-depth manual analysis of 254 MSR publications that used the term "experiment" from

___________________________

- C. Ayala and X. Franch are with Universitat Politècnica de Catalunya, BarcelonaTECH, Campus Nord - Jordi Girona 1-3 Barcelona, Spain. CO 08034. E-mail: {cayala, franch}@essi.upc.edu.
- B. Turhan is with University of Oulu, and Monash University. Pentti Kaiteran katu 1, Linnanmaa, Finland. E-mail: Burak.Turhan@oulu.fi.
- N. Juristo is with Universidad Politécnica de Madrid, Campus de Montegancedo. Boadilla del Monte. Spain. 28660. E-mail: natalia@fi.upm.es.

a representative sample consisting of top-ranked conferences and journals. Our results show that some characteristics of *"experiments"* in MSR research differ from the characteristics traditionally acknowledged in SE experiments. We found that 19% of the assessed MSR studies are not really experiments, but observational studies so they use the term *"experiment"* in a misleading way. From the remaining 81% of the papers, only one of them refers to a controlled experiment while the others stand for experiments with limited control. Such important control nuances have crucial implications on the possibility of the studies to detect cause-effect relationships and therefore on their internal validity and reliability; however, they are mostly overlooked in MSR studies. To support a further understanding and improvement of the design, execution and reporting of MSR experiments, we provide recommendations and guidance to characterize MSR studies from a methodological perspective.

The rest of the paper is organized as follows: Section 2 details the background on experiments' definitions and the characteristics of MSR research. Section 3 details the research strategy and methods followed in this research. Section 4 describes the hallmarks of experiments. Section 5 describes the systematic mapping performed to select the MSR primary studies to be assessed. Section 6 details the criteria to characterize and assess the MSR primary studies. Section 7 provides the assessment on the use of the term experiment in MSR research and summarizes main findings, recommendations and actionable insights. Section 8 discusses threats to validity of this study; and Section 9 summarizes our conclusions.

## 2 BACKGROUND

In this section, we give a brief background on the concept of experiments in both SE and other fields to gain insights on the hallmarks that differentiate an experiment from other types of empirical studies. In addition, we provide a summary of the characteristics of MSR research.

### 2.1 What is an Experiment?

The term experiment refers to an empirical procedure where an *intervention* (called treatment or independent variable) is deliberately introduced to observe its effects on some aspects of the reality (called response variable or dependent variable) under controlled conditions [55]. Experiments are interventional studies aimed to find explanations to cause and effect relationships, so they are usually called explanatory studies [42]. A relevant aspect of interventional studies refers to randomization. It is that the researchers ensure that the study units are allocated to a treatment by a random process.

Experiments differ from observational studies that passively observe the reality, because experiments imply *intervention* and *control* from the researcher. In observational studies, the researchers *do not intervene* in the reality under study [55] and are usually called descriptive studies [18] as the researcher merely documents their observations without trying to alter the course of natural events.

### 2.1.1 Experiments in Software Engineering

In the empirical SE literature, the term experiment is defined as: Wohlin et al, state "*an empirical enquiry that manipulates one factor or variable of the studied setting. Based on randomization, different treatments are applied to or by different subjects, while keeping other variables constant, and measuring the effects on outcome variables.*" [66]. Juristo and Moreno, define experiment as "*an empirical procedure where key variables of a reality are manipulated to investigate the impact of such variations [...]. It is rooted in detecting quantifiable changes as a means of comparing one unitary experiment with another in search of the difference between them and, hence, the reason for the changes*" [28]. Sjoberg et al state that "*[a] controlled experiment in software engineering is a randomized or quasi-experiment, in which individuals or teams (the study units) conduct one or more software engineering tasks for the sake of comparing different populations, processes, methods, techniques, languages or tools (the treatments)*" [58]. Note that in SE, the general terms experiment and controlled experiment are often used synonymously [5], [27], [52], [66], [67] evidencing that other types of experiments had not been approached further in SE [66], [67], [73]. However, emerging opportunities for empirical research in SE such as MSR experiments or the incipient use of field experiments [65] are urging to reconsider such limited familiarity with the diversity of experiments. The SE community should understand the characteristics of the incoming types of experiments and accommodate them into the SE empirical methods.

### 2.1.2 Diversity of Experiments

To understand the diversity of experiments, the literature in other more consolidated scientific areas has characterized them based on their degree of randomization and control [9], [42]. Table 1 summarizes randomization and control characteristics of the different types of experiments and observational studies and the effect of such characteristics on the possibility to reach causality from these studies.

TABLE 1
SUMMARY OF MAIN CHARACTERISTICS OF DIFFERENT TYPES OF EXPERIMENTS AND OBSERVATIONAL STUDIES

| | Strict Control | Limited Control (with similar comparison conditions across groups) | No Control (dissimilar comparison conditions across groups) |
|---|---|---|---|
| With Randomization | *Laboratory Controlled Experiments* Causality detection: High | *Field Experiments* Causality detection: Low* | N/A |
| Without Randomization | *Quasi-Experiments* Causality detection: Medium* | *Natural Experiments* Causality detection: Very Low* | *Observational Studies* Causality detection: N/A |

\* There exist some alternatives to help assess evidence of causation in studies for which strict control may not be a feasible option [47], [74]. A relevant example is the Bradford Hill criteria for establishing epidemiologic evidence of a causal relationship between a presumed cause and an observed effect, that is widely used in public health research [47].

Please note that the availability and evolution of computational power has led also to another type of experiments namely *in silico experiments* or *computational experiments*. They refer to experiments performed on computers or via computer simulation and have become a crucial experimental complement in several areas of research [14], [15], [73]. However, *in silico* experiments are not explicitly included as experiments in Table 1 as they are a kind of secondary study that require previous empirical knowledge for modeling the phenomena or behavior to be studied. It is, the observed causality depends on the validity of the model. So, causality from *in silico* experiments should be assessed and interpreted with caution in the context of each specific application area [3], [36], [55].

**With randomization and strict control.** These studies are commonly called *laboratory-controlled experiments*. They are performed in artificial environments that allow strict controlled conditions (as opposed to the real world, where the conditions cannot be controlled at will). This kind of experiment has been traditionally considered the standard way to study cause-effect relationships between treatments and response variables, because they explicitly control all the potential influences on the response variables and are less threatened by experimental error and bias [24], [42], [55], [59]. Unfortunately, the level of control and artificiality decreases the external validity. Indeed, laboratory experiments must be reproducible and their results must be replicated in other laboratories for the new knowledge to be considered valid [28]. Laboratory controlled experiments are common in physics, chemistry, and biology as it is possible to afford such controlled environments. In SE, a typical example of controlled experiment is when the researcher intervenes by ensuring randomization and a controlled environment to compare a control group against a treatment group when performing a SE related task. All variables in this setting are deemed identical between the two groups except for the treatment being evaluated. Concrete examples can be found at [34] and [58].

**With randomization and limited control.** Although having randomization, in these studies, the phenomenon is observed in the real world (i.e., in naturally occurring environments) with limited control opportunities. They are commonly called *field experiments*. The observation of the phenomenon in its real environment enables field experiments to have higher external validity than laboratory experiments at the cost of lower internal validity, because of its limited control possibilities. Field experiments are the most popular form of experiments in agriculture. The most typical example is an experiment that compares the effect of fertilizers on a soil. The researcher intervenes by randomly applying the treatment (i.e., the fertilizer) on the soil but there is limited control over the weather or other possible sources of impact on the response variable. Therefore, causality cannot be as clearly detected as in laboratory-controlled experiments. Publications on field experiments are rare in SE as the chances of having further access to industrial settings to randomly choose individuals or teams conducting one or more software engineering tasks for the sake of comparing different treatments, are quite elusive so far. A relevant exception is: [65] that performed a field experiment to assess the influence of the English lingua franca mandate on the teamwork in a non-English speaking software outsourcing vendor.

**Without randomization and strict control.** These studies lack of randomization as the allocation is determined by nature or by other situations outside the control of the researchers. When the level of control is strict and specific control strategies amend the lack of randomization, these studies are called *quasi-experiments*. The lack of random allocation of treatments to the experimental units poses internal validity concerns on quasi-experiments and restraints their ability to envisage causality. Quasi-experiments are commonly used in social sciences, psychology, public health, education, and policy analysis, where practical or ethical issues prevent the allocation of treatments to study units. For example, in SE, the costs of teaching professionals all the treatment conditions (different technologies), so that they can apply them in a meaningful way, may be prohibitive and expensive. Therefore, it is common to perform quasi-experiments, including professionals already familiar with the technologies [31] (i.e., the treatment conditions are not randomly applied to the subjects but come intrinsically with them).

**Without randomization and limited control.** These studies lack random allocation of treatments to experimental units (as the occurrence of the phenomenon use to be determined by nature). The intervention of the researcher consists on procuring a setting that ensures an adequate level of control so that the treatment and control groups are comparable, hence the observations of the changes after the occurrence of the phenomenon are indeed due to the studied factors [54]. These studies are usually called *natural experiments*. Natural experiments are highly threatened by (experimental) errors and bias and their boundaries with respect to observational studies are quite elusive. Actually, there is still a permanent source of conflict in the literature, since some literature consider natural experiments as observational studies, while others consider that natural experiments can also achieve some level of causality [16], [47]. The subtle difference between a natural experiment and an observational study is that the former includes researcher intervention to ensure a proper comparison of similar conditions, but the latter does not [60]. For instance, a well-known natural experiment refers to the study of the effects on people's health of smoking banning from all public places in Helena, Montana for a period of six months, which was then compared to a similar period without the ban. The researcher intervention and control mechanisms were aimed to manipulate the reality in order to ensure comparable conditions, in this case similar periods with and without the smoking ban. Natural experiments are widely used in epidemiology, social and behavioral sciences. To the best of our knowledge, no explicit attempts have been done so far to approach natural experiments in SE.

We can conclude that the context and needs of the scientific disciplines determine their types of empirical studies. In scientific disciplines like physics, chemistry, and biology, laboratory-controlled experiments are the dominant type of empirical study for acquiring knowledge. In other sciences such as astronomy, geology, ecology, or paleontology, natural experiments and observational studies are more common as the occurrence of studied phenomena is mainly determined by nature [41]. For instance, researchers cannot create a solar eclipse, they need to wait for it. Special cases such as medicine have incorporated a great diversity of empirical studies and have achieved high consolidation among all types of experiments [50].

## 2.2 MSR Research

The first workshop on MSR was held in 2004. Its success has continued to attract new researchers, ideas, and applications [17] [22], [23], [49]. Nowadays, MSR research has become an important area of SE research and a popular tool for empirical studies in SE, which is evidenced by the related publications and special issues in main SE venues such as ICSE, IEEE Software, and International Journal of Empirical Software Engineering (EMSE) [17], [38], [49], [68], [70].

Some commonly explored areas include software evolution, models of software development processes, characterization of developers and their activities, prediction of future software qualities, use of machine learning techniques on software project data, software defect prediction, analysis of software change patterns, and analysis of code clones [49].

### 2.2.1 MSR Process

As Jung et al. state [68], MSR is defined as "*the process of automatically discovering useful information in large data repositories*". One of the most important characteristics of MSR is that software domain knowledge is required for the analysis of data, because the sources mainly come from code files, bug reports, design documents or other special kinds of development related archives. Extracting and processing these data are not easy without SE domain knowledge and cannot be understood just with statistics [68].

Xie et al [69] describe the general process of MSR with 5 steps. Step 1 and 2 stand for the process of determining the SE task to support and collect/investigate data about it. Step 3 refers to preprocessing data; it involves first extracting relevant data from the raw SE data. To perform the extraction process from SE repositories, some researchers develop APIs and tools. After the extraction, the raw data obtained are processed by cleaning and properly formatting for the subsequent step. For instance, some text-based data requires tokenization, removal of stop-words and stemming before they are used. Step 4 refers to the application of diverse techniques from different aspects of data management and data analysis, including pattern recognition, machine learning, statistics, information retrieval, concept and text analysis [68] in order to produce an optimal output, based on the mining requirements derived in the first two steps. The final step transforms the output results into an appropriate format required to be applied and assist the SE task.

### 2.2.2 Existing MSR Literature Reviews

Several reviews of the literature on MSR have been published. For instance, De Farias et al [71], Hemmati et al [23], Demeyer et al [70], or Halkidi et al [20].

One of the first works on surveying MSR literature was provided by Kagdi et al. [29]. They surveyed the MSR literature on software evolution and provided a taxonomy of MSR research. Their taxonomy enclosed four dimensions: software evolution (layer 1), purpose (layer 2), representation (layer 3), and information sources (layer 4).

Halkidi et al [20] surveyed the mining approaches that have been used so far in SE and categorized them according to the corresponding parts of the software development process they assist.

Hemmati et al [23], reports best practices that the MSR community has developed over the period 2003-2013 and creates a working cookbook (a set of best practices) that can be continuously used and updated as the MSR community matures and advances.

Demeyer et al [70], presented a text mining exercise applied on the complete set of papers from the MSR conference to identify how the research on MSR has evolved. The study describes an automatic and quantitative approach in order to identify issues like trendy research topics, the most frequently (and less frequently) cited papers, and the most popular mining infrastructure.

De Farias et al. [71] did a systematic mapping study for collecting evidence on software analysis goals, data sources, evaluation methods and tools used in MSR, in order to understand how this area had been evolving. Although they classified the evaluation methods (i.e., survey, case study, controlled experiment, …), such classification was mainly based on what was stated in the paper rather than a further assessment of the appropriateness of the empirical approach used.

Although all these works are valuable, to the best of our knowledge, none of these reviews on MSR literature focuses on analyzing the characteristics and methodological needs of MSR experiments nor the proper use of empirical terminology.

## 3 RESEARCH METHOD

We carry out a flexible exploratory research approach to answer two research questions.

**RQ1: What are the hallmarks of experiments?**

To answer this question, we performed an in-depth study of experiment's definitions provided in SE related literature [5], [28], [66], and other more consolidated experimental fields to understand the characteristics that experiments must fulfill [1], [10], [11], [16], [19], [24], [35], [37], [41], [42], [52], [54], [55], [56], [60]. We specially studied medicine literature [9], [18], [40], [47], [59], [63] as it is a consolidated discipline that exploits all types of experiments and other types of empirical studies, and unify these studies into a well-founded and recognized operational classification [50]. We held several brainstorming meetings and finally got a consolidated view of the characteristics of experiments that differentiate them from other types of empirical studies. The answer to this RQ has been detailed in Section 4.

**RQ2: Is the term "Experiment" properly used in MSR research?**

To assess the use of the term *experiment* in MSR research, we performed a small-scale systematic mapping study (SSSM) aimed to gather evidence on the coverage of the experiment's hallmarks of the MSR primary studies that used such term (details on the systematic mapping are presented in Section 5). To support the assessment of the coverage of the experiments' hallmarks by MSR primary studies, we elaborated a set of criteria that helped us to characterize MSR experiments (such criteria and their associated results are presented in Section 6).

Fig. 1 summarizes the research approach followed in this study. This approach allowed us first understand the hallmarks of experiments (RQ1) and then to analyze such hallmarks in the context of MSR studies obtained from the SSSM (RQ2). We intertwined the analysis and discussion of the primary studies to envisage and calibrate the criteria for assessing the use of the term "*experiment*" in MSR research. Thus, in Section 7, we detail the results together with actionable recommendations and insights.

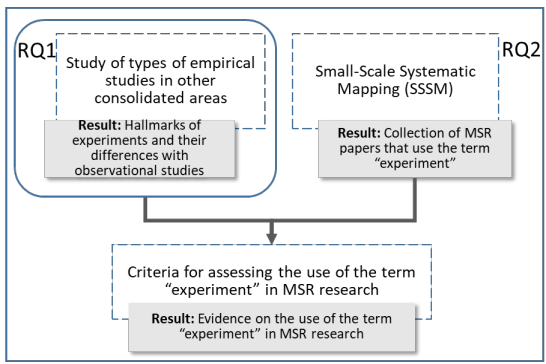

Fig. 1. Overview of the research approach followed in this work.

Although diverse sciences and disciplines have adapted the experimental procedure to their study of phenomena of interest, there are certain hallmarks that laboratory-controlled experiments must fulfill to be considered as an interventional study able to detect causality. In this section, we summarize controlled experiment hallmarks and their impact on the nuances of the different kinds of experiments. By hallmark, we refer to those characteristic that laboratory-controlled experiments must satisfy but other empirical studies do not. These characteristics are manipulation, control, and randomization:

– **Manipulation:** Experiments require that the researcher deliberately intervene on the reality under study to observe the effects of such intervention (i.e., the researchers arrange the world before observing it).

– **Control:** In order to guarantee that the changes observed in the studied phenomenon are only due to the treatment under study, it is required that extraneous variables are controlled. The degree of internal validity of an experiment depends on its level of control. Note that control is not an all or nothing condition; the degree of control varies in a continuous scale [58]. The more control of undesired variables affecting the response variable, the higher chances to identify cause-effect relationships between treatment and response variable [1], [47], [52].

– **Randomization:** It is a specific procedure in which study units are assigned to receive the treatment or an alternative condition by a random process. Note that randomization can also be applied for sample selection in order to increase external validity in any type of empirical study. However, while random sampling selection is a desirable condition, the random assignment of treatments to units is a must for controlled experiments, as it is the default control to prevent bias and to identify causality [59]. Therefore, in this paper, by *randomization* we refer to this essential condition for controlled experiments.

Table 2 summarizes the characteristics that differentiate the diverse types of experiments from observational studies, and their influence on the possibility to detect causality. Notice that manipulation is necessary and that causality emerges from the degree of control and randomization applied by the researcher during experimental design.

As observed in Table 2, quasi-experiments and natural experiments lack proper randomization conditions. However, some thoughtfully chosen control strategies must be put in place as amendments to reduce the plausibility of internal validity threats and reaching the hallmarks of experiments.

TABLE 2.
CHARACTERISTICS THAT DIFFERENTIATE EXPERIMENTAL AND OBSERVATIONAL STUDIES

| Type of Study | | Hallmarks of Experiments | | | Causality | Int. validity | Ext. validity |
|---|---|---|---|---|---|---|---|
| | | Manipulation | Control | Randomization | | | |
| Interventional Studies | Laboratory Controlled Experiment | ✔ | Strict | ✔ | High | High | Low |
| | Quasi-Experiment | ✔ | Strict | Amended by the researcher | Medium | Medium | Low |
| | Field Experiment | ✔ | Limited | ✔ | Low | Medium | High |
| | Natural Experiment | ✔ | Limited | Determined by nature | Very Low | Low | Medium |
| Observational studies | | ✘ | ✘ | ✘ | ✘ | * | * |

✔ must have, ✘ do not have, * Depends on the study design

Some common special amendments for quasi-experiments are [11], [56] [31]: a) repeated measures design, enabling each subject to be its own control; b) pretest scores to control for pre-experimental differences between experimental groups; c) use several experimental groups for some or each treatment condition to allow comparison of effects of different types of groups.

For natural experiments, the amendment consists on ensuring that treatment and control groups are similar in terms of all observed and unobserved factors that may affect the outcome of interest, with the exception of the treatment and confounders that the researcher controls for [60].

## 5 SMALL-SCALE SYSTEMATIC MAPPING STUDY

To collect and assess evidence on the use of the term "*experiment*" in MSR research we performed a small-scale systematic mapping study (SSSM). Mapping studies are a means of evaluating the state of research in a specific area [6]. We followed the guidelines for systematic literature reviews proposed by Kitchenham et al. [33]. However, in the searching process, we approached a small set of representative venues rather than a more comprehensive one as expected by systematic reviews. In addition, we did not assess the research quality and evidence from the primary studies (as suggested by systematic reviews); instead, we developed our own criteria for assessing the use of the term "*experiment*". The following subsections details the process and the corresponding results. The search process, evaluation and classification of the results was manual.

## 5.1 Search and Refinement Process

### 5.1.1 Stage 1. Defining Search Sources and Universe of Papers

Our objective is to get a deep understanding on the use/misuse of the term "*experiment*" in MSR research. According to this, in order to further analyze the use of the term "*experiment*", we considered appropriate to work with a representative sample of venues rather than with the complete MSR papers population, as it would be necessary if our goal was to obtain a map. From a practical point of view, the type of investigation we need to conduct on every selected paper prevents the analysis of a very large set of papers.

We selected three relevant venues that focus on publishing empirical works in general and MSR research in particular: the Working Conference on Mining Software Repositories (MSRConf), International Journal of Empirical Software Engineering (EMSE), and ACM/IEEE International Symposium on Empirical Software Engineering and Measurement (ESEM). We limited our study to the assessment of papers published in the year previous to the beginning of our research (2015) with the aim to profoundly analyze each primary study.

Regardless of the specific topic of the papers published in the selected venues, we aimed to assess papers that used the term "*experiment*" to identify their research methods. We gathered all publications from the selected venues using digital libraries to retrieve the pdf versions of all papers. We got 822 papers in total that correspond to all publications from the selected venues from 2015 to 2018. In the case of MSR, we decided to deter papers that belonged to two special tracks namely "data show case" and "mining challenge", because these papers focused on sharing data or promoting the use of mining tools, respectively, instead of conducting empirical studies. As a result, we removed 98 papers and ended up with 724 papers.

### 5.1.2 Stage 2. Selecting Primary Studies

To identify publications about conducting MSR experiments using repository data (e.g., commits, bugs, code reviews, …); we manually reviewed the 724 papers from the previous stage. We discarded papers that did not contain the term "*experiment*" or did not conduct experiments or did not use repository data as experimental units. We first searched for the term "*experiment*" in the full text. If the term "*experiment*" was used, we skimmed the full text to decide if the paper conducted an experiment or just con-tain the word "*experiment*" for other purposes, and to confirm that the paper used repository data as experimental units. In most of the cases, the rejected papers did not contain the term "*experiment*" or referred to other methodological approaches such as case studies, interviews or systematic literature reviews. Other rejected papers contained the term "*experiment*" but referred to human-based experiments or to practical support or guidelines to conduct experiments instead of conducting experiments. As a result, we selected 254 primary studies conducting MSR experiments (i.e., studies that claim to perform an experiment and use repository data as experimental units). Some of the primary studies, in addition to performing an experiment using repository data as experimental units, also reported human-based experiments. In these papers, our assessment was limited to the experiments using repository data. Table 3 provides an overview of the results of stages 1 and 2. Notice from Table 3 that several publications in our systematic mapping refer to MSR experiments. It evidences the momentum of the research field.

## 5.2 Data Extraction

We extracted the following information from primary studies:

– General information about the publication: title, abstract, authors, affiliation and venue.

– Instantiation of the set of criteria for assessing the suitability of the use of the term "*experiment*". Although attributes or criteria for data collection in systematic reviews should ideally be determined prior to the review [33], our experience, like other researchers have also stated [31], was that our understanding and determination of which attributes to use for assessing the primary studies resulted in an intertwined process of reading a high percentage of the primary studies in order to stabilize the understanding of the criteria to be used. We conducted a dual-reviewer process on approximately 45% of the articles in order to get a comprehensive set of criteria that was used afterwards to collect information from all primary studies. The resulting criteria and the number of papers that covered such criteria are discussed in Section 6.

Following the recommendations by Brereton et al [7], one researcher extracted the data, while other randomly choose some papers to confirm the extracted data. Any differences between the reviewers were solved through discussions until a consensus was reached. We verified whether or not the publication mentioned or discussed issues related to each of the criteria, and registered some notes when necessary to better discuss about them. Notice that we did not assess the research rigor, we just focused on getting insights on the papers' coverage to the hallmarks of experiments. The primary studies included in this study are detailed at the end of the References section. Details of the SSSM and the extracted data for each single study can be consulted here:

https://www.essi.upc.edu/~cayala/Suplemental-Material-TSE-Ayala-etal2021.xlsx

TABLE 3
OVERVIEW OF STAGES 1 AND 2 OF THE SSSM

| Venue | Total number of publications per year | | | | | Stage 1. Universe of Papers | | | | | Stage 2. Primary Studies | | | | |
|---|---|---|---|---|---|---|---|---|---|---|---|---|---|---|---|
| | 2015 | 2016 | 2017 | 2018 | Total | 2015 | 2016 | 2017 | 2018 | Total | 2015 | 2016 | 2017 | 2018 | Total |
| EMSE | 55 | 68 | 90 | 97 | *310* | 55 | 68 | 90 | 97 | *310* | 20 | 36 | 33 | 43 | *132* |
| ESEM | 63 | 56 | 61 | 58 | *238* | 63 | 56 | 61 | 58 | *238* | 8 | 12 | 17 | 11 | *48* |
| MSRConf | 72 | 59 | 66 | 77 | *274* | 42 | 42 | 44 | 48 | *176* | 18 | 22 | 14 | 20 | *74* |
| Total | 190 | 183 | 217 | 232 | ***822*** | 160 | 166 | 195 | 203 | ***724*** | 46 | 70 | 64 | 74 | **254** |

TABLE 4
CRITERIA AND RESULTS FOR ASSESSING THE SUITABILITY OF THE USE OF THE TERM EXPERIMENT FROM THE PRIMARY STUDIES

| | Criteria for assessing MSR studies | Identified categories or values from the MSR studies | Questions for assessing the coverage of primary studies | No. Papers |
|---|---|---|---|---|
| **Hallmarks of Experiments** | | | | |
| 1 | Manipulation | Observational | Does the study focus on finding trends or observations without manipulation of the studied phenomenon? | 47 |
| | | Interventional | Does the researcher intervene in the study design? | 207 |
| 2 | Control | Use of retrospective repositories | Does the study use retrospective repository data? | 253 |
| | | Use of prospective repositories | Does the study use prospective repository data? | 1 |
| | | Use of datasets blocking/repetition strategies | Does the study use blocking or repetition strategies? | 206 |
| 3 | Randomization | Random allocation of treatments to experimental units | Does the study randomize the allocation of treatments to datasets? | 0 |
| | | Datasets randomization | Does the study randomly select repositories or datasets from them? | 79 |
| **Other Relevant Characteristics** | | | | |
| 4 | Types of MSR Interventional Studies | Studies based on Comparisons | Does the interventional study focus on comparing the behavior of different approaches under the same experimental conditions and datasets? | 112 |
| | | Studies based on Training Machine-Learning Algorithms | Does the interventional study focus on training a machine-learning algorithm on the datasets? | 95 |
| 5 | Focused Scope | | | |
| | Hypothesis definition | Statement of an explicit hypothesis | Does the study explicitly state a hypothesis to be tested? | 95 |
| | Limitations definition | Discussion of Limitations or threats to validity | Does the study discuss limitations or threats to validity? | 234 |
| | Use of single repositories | Single/Multiple repositories | Does the study use a single repository? | 36 |
| 6 | Statistical Analysis | Use of statistics | Does the study use statistical analysis? | 250 |
| 7 | Replication facilities | Provision of material for replication | Does the study provide a replication package? | 111 |
| 8 | Causality | Use of causality related terms. | Does the study use causality terms to discuss its findings? | 32 |

# 6. CRITERIA FOR ASSESSING THE USE OF THE TERM "EXPERIMENT" IN MSR RESEARCH

The resulting criteria for assessing the use of the term "*experiment*" include both characteristics that relate with the experiment hallmarks as well as other relevant characteristics that could positively influence validity threats and serve to counterbalance the coverage of some studies to the experiment's hallmarks.

Table 4 shows the aggregation of the resulting criteria detailed in the following subsections, together with the number of papers that satisfied such criteria (as shown in the last column of Table 4). To ease the analysis of each primary study, we added specific questions to each criterion. Responses for each criterion contain binary values (yes/no) and we used additional notes, if necessary, for making our discussions easier. For cases where the identified categories were devised as mutually exclusive (i.e., Observational/Interventional, Prospective/Retrospective, and Studies based on comparisons/Studies based on Training ML algorithms) but the study approached both categories, we categorized the study into the category that was dominant in the paper.

## 6.1 Manipulation

As in any other empirical discipline, the manipulation hallmark, make the most relevant distinction between observational and interventional studies [42]. In particular, in medicine, the level of manipulation of a study is determined by its investigative purpose and the role of the researcher in the study [61]. We identified MSR studies without manipulation (observational) and with manipulation (interventional).

**Without Manipulation (Observational studies).** In these studies, the researcher neither control nor intervene in the studied phenomenon, but instead observes natural relationships between factors and outcomes. Their investigative purpose is to describe and uncover associations and patterns without regard to causal relationships. For instance, a study where a set of developers are classified as light, moderate or heavy social media users, and correlated with the quality of their software code documentation. In such a study, researchers are neither intervening on developers' social habits nor controlling their software code documentation (for instance, the way it is generated, the amount of documentation produced, the programming language used, and so on). Instead, developers' behavior and their software code documentation are set free. This type of studies is observational and some authors refer to them as *"association analysis"* [29] or *"correlational studies"* [39]. 47 out of 254 MSR primary studies are observational but were wrongly labeled as experiments.

**With Manipulation (Interventional studies)** are those studies where the researcher intervenes as part of the study design. The investigative purpose here is to identify the extent and nature of cause and effect relationships. 207 out of 254 primary MSR studies have an interventional purpose. For instance, Ryu et al. [EMSE2016-3] used NASA and SOFTLAB datasets to develop prediction models for cross-project defect prediction. Martie and van der Hoek [MSR2015-8] executed a study to compare the performance of four novel algorithms for ranking code search results with other three well-known algorithms, using data from 300,000 projects from GitHub. Note that researcher intervention implies the preparation of a suitable experimental setting for running a test or a series of test over dataset(s) to observe the effects of factors that affect the output (usually a model) with a specific goal in mind. The factors can be the algorithm used to generate the output, the hyper parameters of the algorithm, the datasets, etc. Menzies and Shepperd [39] call this type of studies *"computational experiments"*.

Table 5 summarizes the types of MSR primary studies according to the manipulation hallmark by venue. Most of the venues have over 80% of papers implying Interventional studies, evidencing the prominence of this type of MSR studies considered as experiments.

TABLE 5.
OBSERVATIONAL AND INTERVENTIONAL STUDIES BY VENUE

| Primary Studies' Manipulation | | EMSE | ESEM | MSRConf | Total |
|---|---|---|---|---|---|
| **Observational** | | 27 (20%) | 6 (13%) | 14 (19%) | **47** |
| **Interventional** | | 105 (80%) | 42 (87%) | 60 (81%) | **207** |
| *Total papers per venue* | | **132** | **48** | **74** | **254** |

## 6.2 Control

Contrary to experiments involving physical objects, the use of repository data and the computational nature of MSR experiments calls for specific control mechanisms in order to avoid biased results due to confounding variables. This has been recognized in computational-based disciplines such as machine learning [3]. Machine learning's experimental process has been acknowledged to be especially good for experimentation in the sense that "*as opposite to the natural sciences where one can never control all possible variables […], machine learning can avoid such complications*" having complete control over the settings used for its studies, making systematic experimentation easy and profitable [36]. Control mechanisms for computational-based experiments are:

**Dataset Control strategies.** The researcher should design and execute strategies to ensure the manipulation of datasets under unbiased conditions. The basic ones in MSR are [3]:

- **Dataset Repetition:** To repeat the experiments/trials multiple times to average over the effects of uncontrollable variables such as the noise in the datasets or other factors affecting the behavior of the algorithms. It allows to obtain an estimate of the experimental error.

- **Dataset Blocking:** To reduce the variability due to nuisance factors that influence the response variable by blocking an aspect of the experimental setting in all trials in order to ensure that the observed differences are due to the influence of the response variable.

Our results show that 206 out of 254 MSR primary studies explicitly state any of these dataset control strategies.

**Dataset Measurement and Collection Control.** The researcher should design and execute strategies for measuring and collecting the datasets in a controlled environment. This guarantees that the data stand for reliable evidence of the studied phenomenon. The literature on machine learning do not usually tackle in deep this aspect as it is quite specific of the application domain [36]. Inspired in medical research, we found that the level of measurement and collection control of the data of a study can be influenced by the role that time plays in data collection, either prospective or retrospective [61]:

- **Prospective studies** follow participants forward through time, collecting data in the process and recording it into prospective repositories. These studies provide the highest data collection control and are less prone to some types of bias. As a result, these studies can more strongly suggest causation [61] as the researcher is able to control extraneous variables and decide on the metrics to gather as well as the allocation of treatments to units. *The use of prospective repositories in MSR studies is exceptional.* The only case from the 254 MSR primary studies, which uses a prospective repository, is Rashid et al. [ESEM2015-4]. They set up a specific infrastructure to run and collect some predefined and controlled events for measuring the energy consumption of different sorting algorithms implemented in different programming languages.

- **Retrospective studies** are those where data are collected from the past and recorded into retrospective repositories, either through records created at that time or by asking participants to remember the studied phenomenon [61]. Retrospective studies are more prone to different biases, as the researcher has no chance of intervention or control over the recorded phenomenon. Therefore, the researcher relies on others for accurate record keeping and data availability. 253 MSR studies out of 254 used retrospective repositories. *The vast majority of MSR primary studies used retrospective repositories coming from open source software repositories and to lesser extent organizational repositories of data, or even data from systematic literature reviews.* For instance: Shahbazian et al. [MSR2018-39] uses data from 301 versions of five large open-source systems to build a predictive model that is able to identify the architectural significance of newly submitted issues. Minku et al. [ESEM2015-14] used data on 125 Web projects from eight different companies part of the Tukutuku database to build web effort estimation prediction models. Yu et al. [EMSE2018-12] generated a dataset from existing SE literature reviews in order to evaluate a technique for studying a large corpus of documents based on active learning algorithms.

*None of the MSR primary studies explicitly tackled the control issues associated with the type of repository used.*

### 6.3 Randomization

We recorded whether the studies cover the random assignment of treatments to experimental groups, as it is an essential condition for controlled experiments to prevent bias and be able to deduce causality [59]. We found that none of the MSR primary studies performs random assignment of treatments to experimental units. We thus realized that randomization in MSR studies seem to have a slightly different meaning than traditional experiments, and resembles to the meaning from the machine-learning discipline [3]. *In MSR studies, randomization means **Dataset Randomization** and refers to the application of random strategies for selecting datasets from the used repository.* Dataset Randomization has a great impact on the internal validity of the experiments. Please note that randomization in MSR can also be applied for repository selection but it is not hallmark and only affects the external validity of the study.

To get a throughout understanding of randomization in MSR studies, we gathered whether datasets randomization and/or random selection of repositories were used. *Despite datasets randomization is a hallmark for MSR experiments only 79 out of 254 MSR primary studies detailed it.* Regarding repositories selection, it seems based on convenience, availability, or because they have been used in earlier research and provide readily available datasets, as other literature has also highlighted [4], [12], [39], [EMSE2018-83].

### 6.4 MSR Interventional Studies Design Types

Broadly speaking, we distinguish two main goals of MSR Interventional studies: *studies based on Comparisons* and *studies based on Training Machine-Learning Algorithms*. The nature of the goal of these study types requires different experimental design strategies with regard to internal validity. Fig. 2 provides an overview of the differences between them.

**Studies based on Comparisons.** The focus of the experimental process in this type of studies is on comparing the behavior of different approaches (the independent variables) under the same experimental conditions and datasets (randomly selected) in order to obtain their corresponding results and compare among them.

Fig. 2a) depicts a high-level overview of this type of experiments. For example, Martínez et al, [EMSE2017-31] perform an experiment to compare the effectiveness of some state-of-the art approaches for automatic test-suite repair (i.e., the independent variables) using a well-known dataset called Defects4J. Regardless of other experimental design issues, this type of experiments should guarantee an identical environment for enabling the comparison among the studied approaches. Therefore, the use of *datasets randomization* and *dataset blocking* is necessary to ensure the internal validity of the study. In other words, if we are comparing the behavior of different approaches, they should all use the same randomly obtained datasets, otherwise the differences in the corresponding observations would depend not only on the approaches but also on the different datasets.

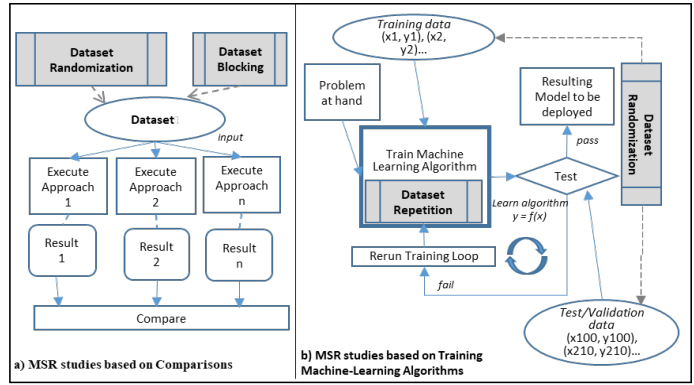

Fig. 2. Overview of types of MSR Interventional studies.

**Studies based on Training Machine-Learning Algorithms.** The focus of the experimental process in this type of studies is on training an algorithm based on repository data in order to build an optimal resulting model for a problem at hand. The independent variable is usually a feature from the input dataset(s) and the dependent variable is the estimated performance of the resulting model. To do so, these studies require not only *datasets randomization*, but also that each dataset used is randomly split so that some part of the original dataset is used for *training* while another part is used for *testing* or validation [3]. It is important to remark that such random split is done at the repository level in order to preserve the reliability of the data from each single repository. The training process should be re-run until the test step is passed and the resulting model is ready for deployment. To guarantee to average over the effect of uncontrollable variables such as the noise in the dataset or other factors affecting the algorithm, these studies require to repeat the trials multiple times (i.e., *dataset repetition*) [3].

Fig. 2b) depicts a high-level overview of this type of studies. For instance, Li et al [EMSE2017-40] used random forest classification algorithms over datasets from four open source projects' repositories (Hadoop, Directory Server, Commons HttpClient, and Qpid) to develop a model for providing log changes suggestions to software developers. Regardless of other experimental design issues, this type of experiments should guarantee a controlled setting for the training process. Therefore, the use of *datasets randomization, random selection of training and testing datasets* and *dataset repetition* is necessary to ensure the internal validity of the study.

### 6.5 Focused Scope

Experiments are purposeful studies designed for testing explanations [41]. The experiment's scope needs to be defined a priori in order decide on the controls to make with the aim to increase the chances to identify causal relationships. The definition of a focused experiment's goal is usually denoted by a hypothesis to be tested. We checked the existence of at least an explicit hypothesis that reinforces the focused nature of interventional studies. Also, we considered pertinent to look at the discussion of limitations or threats to validity in the studies. These aspects provide a practical insight on the focused scope required by experiments. *Although our results show that including limitations or threats to validity information is a widely recognized reporting practice in MSR research (234 out of 254 MSR primary studies included this aspect), only 95 out of 254 primary studies stated an explicit hypothesis to be tested.* So, **considerable room for MSR researchers to improve the description of the scope of their studies and thus better defining the external validity of their studies exists** [MSR2016-9]. In addition, as the use of multiple repositories might help to widen the scope of the MSR studies' results [12], we gathered whether single or multiple repositories were used. *Our results show that only 14% of the primary studies (36 out of 254) used data from a single repository while 86% used data from multiple repositories.*

## 6.6 Statistical Analysis

The use of statistics has been usually associated to experiments, especially because statistical hypothesis testing is used to determine whether an experiment conducted provides enough evidence to reject a proposition. However, any other type of empirical study can use statistics to analyze and present its collected data, for instance, survey research usually uses statistics to analyze and present its data [42]. The traditional role of statistics to support researchers to remove the chance process in an experiment and establish its validity is practically inherent in MSR studies [2]. It is because MSR studies shift the traditional experiment focus on looking at observations from individual cases to looking at a collection of huge amounts of observations from repository data, where statistical tools should be used to adjust for potentially confounding effects and to interpret the findings. In line with this, *we found that more than 98% of MSR primary studies usually include a statistical analysis element.* Nevertheless, we did not verify that the statistical analysis matches the experiment design. At this respect, Neto et al [45] provide evidence on the wide use of statistical analysis in SE and the high rates of inappropriate usage.

## 6.7 Replication facilities

Replication is at the heart of the experimental paradigm and is considered the cornerstone of scientific knowledge [28]. Experiments need replication at other times and under similar conditions before they can produce an established piece of knowledge [10]. If an experiment is not replicated, there is no way to distinguish whether results were produced by chance [28]. To get insights on the strength of the evidence from the studies, we recorded whether the studies provide replication material. *Surprisingly, although replication has been highlighted as an important aspect of data mining [13], 111 out of 254 (44%) MSR primary studies provided replication facilities.*

## 6.8 Causality

Causality from *in silico* experiments such as those from machine learning, is a controversial topic that should be interpreted with caution [3], [36]. The literature suggests that *in silico* experiments cannot deduce causal relationships per se but hypothesis on causal relationships based on their corresponding models of the phenomena or behavior to be studied [3], [36], [55]. Although the computational disciplines that support the development of and experimentation with such models use genuine experimental procedures [36], the limitations and constrains of *in silico* experiments must be assessed in the context of each application area as they heavily depend on the characteristics of the input data and its interpretation [3]. Therefore, we are performing this study: to provide evidence on the special characteristics of experiments in MSR. To get insights on the notion of causality used in MSR research, we used a subjective but practical approach recording whether the studies used causality related terms such as "our results [imply/corroborate/shows the effect of]" to discuss their findings. We found that 32 out of 254 MSR primary studies use causality related terms. For instance, Scanniello et al. [EMSE2015-8] stated the following causality related sentence: *"The results indicate that correctness and efficiency improve (statistically significant) when developers use our new approach without any impact on the time to accomplish a concept location task."*

## 7. ASSESSMENT ON THE USE OF THE TERM EXPERIMENT IN MSR RESEARCH

In the previous section we detailed that MSR experiments have different connotations with respect to traditional experiments' hallmarks in SE. Although the manipulation hallmark remains the same as in traditional SE experiments, the control and randomization hallmarks vary. Regarding control, MSR experiments refer to both: dataset control strategies and dataset measurement & collection control. Regarding randomization, MSR experiments refer to datasets randomization. In this section, we provide our assessment on the use and misuse of the term experiment in MSR research.

Table 6 provides a summary of the individual coverage of the 254 MSR primary studies to the hallmarks of MSR experiments.

Sections 7.1 and 7.2 detail our results about MSR Observational and MSR Interventional Studies respectively. Then, in Section 7.3 we provide actionable recommendations and insights to improve MSR experiments.

## TABLE 6.
### SUMMARY OF MSR STUDIES' ASSESSMENT CRITERIA

| MSR Studies | Hallmarks of Experiments | | | | | | Other Relevant Characteristics | | | | | |
|---|---|---|---|---|---|---|---|---|---|---|---|---|
| | Manipulation | Control | | | Dataset Randomization | Hypothesis | Focused Scope | Threats | Single Repository | Use of statistics | Provision of replication material | Use causality terms |
| | | Prospective | Retrospective | Blocking/ Repetition | | | | | | | | |
| **Observational Studies** | 47 NO | 0 | 47 | 14 | 11* | 21 | 45 | 8 | 47 | 26 | 3 | |
| **Interventional Studies** | 207 YES | 1 | 206 | 192 | 68 | 74 | 189 | 28 | 203 | 85 | 28 | |
| Based on Comparisons | 112 YES | 1 | 111 | 112 | 41 | 42 | 105 | 14 | 110 | 48 | 13 | |
| Based on Training Machine-Learning Algorithms | 95 YES | 0 | 95 | 80 | 27 | 32 | 84 | 14 | 93 | 37 | 15 | |

*they randomly select dataset(s) from repositories instead of performing datasets randomization

## 7.1 MSR Observational Studies

Our results show that 47 out of 254 MSR primary studies (19%) that used the term experiment are actually *observational studies*. Therefore, these studies *are using the term experiment in a misleading way*. They should not be labelled as experiments as they lack the mandatory manipulation hallmark that is required to identify the extent and nature of cause and effect relationships. Even if some of these studies randomly select their datasets and/or perform some dataset control strategies, they cannot fulfill manipulation and therefore cannot be considered experiments at all.

To understand the severity of misuse of the term experiment by MSR Observational studies, we assessed the characteristics collected for each one of the 47 papers and skimmed again the papers. As a result, we characterized the severity of misuse of the term experiment into 3 categories as shown in Table 7.

## TABLE 7.
### MISUSE OF THE TERM EXPERIMENT BY MSR OBSERVATIONAL STUDIES

| Type of Misuse | Severity of misuse | Observational % | Global % | Description |
|---|---|---|---|---|
| **Occasional misuse** | Light | 24/47 (51%) | 24/254 (9%) | The term experiment is misplaced/ misused once or twice throughout the paper. The study is mostly referred with a generic term such as "empirical study" or "study". No abuse of causality was detected. |
| **Systematic misuse** | Moderate | 20/47- (43%) | 20/254- (8%) | The term experiment is systematically used to refer to the study but no abuse of causality related terms was detected. |
| **Conceptual misuse** | Critical | 3/47 (6%) | 3/254 (2%) | The term experiment is used and abuse of causality hints were detected. |

The set of papers that occasionally misused the term ex-periment (51%) did not seem to compromise a proper understanding of the term as no causality related terms were detected. The use of the term experiment seems to be unintentional (maybe due to a lack of a proper proof-reading of the paper), or that the authors used the term experiment as a verb instead of referring to their research method (e.g., "*we experiment with a threshold…*"). Hence, in these cases, the misuse can be considered light. For the papers that widely used the term experiment throughout the paper and were classified as *systematic misuse* (43%), the severity of the misuse could be considered moderate as the authors also seem to be aware of the type of conclusions that could be drawn from their empirical study (i.e., they do not abuse of causality related terms), but did not get clear insights on proper ways to label their studies. Finally, the severity of misuse of the term experiment might be considered critical in 6% of the papers that were classified as *conceptual misuse* as in these cases, it could be that the concept and/or limitations of the term experiment have not been properly understood as causality seem to be suggested (e.g., "*[X] and [Y] are the mechanism causing Z*".)

Fig. 3 shows the distribution per year of MSR Observational studies based on their type of misuse. It can be observed that the total number of paper that misuse the term experiment is fairly uniform through the assessed years, without any clear trend to decrease/increase.

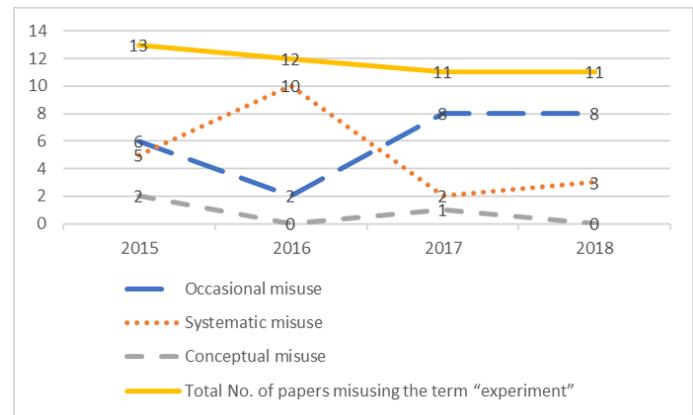

— Occasional misuse
····· Systematic misuse
--- Conceptual misuse
— Total No. of papers misusing the term "experiment"

Fig. 3. Distribution of MSR Observational studies per year.

## 7.2 MSR Interventional Studies

The majority of the MSR primary studies (207 out of 254 MSR primary studies (81%)) are interventional, it is, they cover the manipulation hallmark to be experiments. Nevertheless, from Table 6 we can observe that despite covering the manipulation hallmark, *not all of these MSR Interventional studies properly cover the control and datasets randomization hallmarks to be considered a genuine controlled experiment*.

Regarding data measurement and collection control, we found that *the use of prospective repositories in MSR Interventional studies is anecdotal as all except one of the studies use retrospective repositories*. However, it seems that MSR researchers are not aware of the control limitation produced by the use of retrospective repositories, as they simply use the term "*experiment*" without any explicit limitation state-ment at this respect. Such omission in 99% of the MSR Interventional studies could have negative effects on the reliability of the studies and the correct interpretation of their results.

Regarding the dataset control strategies required by the type of MSR Interventional studies (as described in section 6.4), our results in Table 6 show that all 112 *MSR studies based on Comparisons* report the use of the *datasets blocking* strategy, while 80 out of 95 *MSR studies based on Training Machine–Learning Algorithms* report some kind of *dataset repetition* strategy for ensuring control. This suggests that the importance of *datasets blocking* for ensuring control is fully known and applied by MSR researchers, but *dataset repetition*, although it is widely known, it is overlooked sometimes. Furthermore, regarding *datasets randomization*, 139 out of 207 MSR interventional studies overlook this aspect. In the best case, these overlooked aspects could be just omissions in the report. In the worst case, it could mean that they were not performed in the study, thus greatly compromising the internal validity of the results in at least 67% of the MSR Interventional studies.

Finally, with regard to the purposeful nature of experiments [41], only 74 out 207 (36%) MSR Interventional studies provided an explicit hypothesis. And, regarding causality, 28 out of 207 (13%) MSR Interventional studies use some type of causality related terms, which suggest that MSR authors seem to be cautious about claiming causality from their studies.

## 7.3 Main Recommendations and Insights
Based on our results, below we discuss 6 main recommendations and insights.

**1.- Understanding the distinction between Observational vs Interventional MSR studies is critical for appropriate study design choices in MSR.**

The amount of observational studies (47 out of 254, 19%) that wrongly claimed to be an experiment shows that *the differentiation between observational and interventional studies has not been properly addressed yet in MSR.* Ignoring the differences between observational and interventional studies could have very negative effects on the appropriate choice of study designs [50]. In other words, if MSR researchers do not fully understand such differences, they might likely threaten the selection of proper experimental design choices for their studies.

On the one hand, the term experiment was misused when labeling MSR Observational studies. In most cases, the severity of misuse was not critical, meaning that most researchers did not abuse of causality related terms in their studies, but it was evidenced that MSR researchers struggle to find a proper name for referring to their MSR studies. On the other hand, although MSR Interventional studies could be labelled as experiments (as they fulfill the manipulation hallmark) most of them failed to cover/report the other hallmarks of experiments so they did not used the term properly (see recommendation 2 and 3).

One could also think on the role of reviewers' expertise for detecting misuses, omissions and problems related to the use of the term experiment. For the case of MSR Observational studies, the misuse of the term experiment in most cases was not severe and this could explain why most reviewers did not find it weird and requested no further changes to the papers. For the case of omissions in MSR Interventional studies, we highlight the lack of proper methodological guidelines for MSR research (as stated in recommendation 6 and 7). Such guidelines would support not only authors but also reviewers' tasks.

These situations could be improved by promoting a further understanding of the differences among observational and interventional studies in MSR research as well as the hallmarks of experiments so researchers/reviewers are aware of the characteristics and implications of each type of study. This paper provides insights for clarifying such differences as summarized in recommendation 5.

**2.- Genuine MSR Controlled Experiments require the use of Prospective Repositories.**

We found that *the most overlooked control aspect in MSR Interventional studies is the level of control on datasets measurement and collection* (denoted by the use of retrospective repositories in 99% of the studies). *The use of prospective repositories is the only way to guarantee the maximum level of control required by a genuine controlled experiment* (since the experimenter has control also on the measurement and collection of the datasets). The lack of control on the measurement and collection of retrospective repositories decreases the coverage of the studies to the control hallmark, implying internal validity issues that lead to *experiments with limited control*.

The results of this paper might help MSR researchers to realize the impact of using retrospective repositories on the level of control of their studies and therefore on the type of MSR experiments they perform.

**3.- Overlooking Essential Experimental Design strategies for MSR Experiments highly compromises Internal Validity and Results' Reliability.**

The design and execution of datasets randomization and dataset control strategies discussed in this paper, have been recognized as crucial to ensure an adequate level of control in computational based disciplines [36]. However, at least 67% of the MSR Interventional studies fail to report/apply such aspects.

We hope our results help MSR researchers to realize the relevance of performing/reporting *datasets randomization* and suitable dataset control strategies. Remember that in addition to datasets randomization, S*tudies Based on Comparisons* require datasets blocking; while *studies based on Training Machine-Learning Algorithms* require *random selection of training/test datasets* and *dataset repetition* as default datasets control strategies. Overlooking these aspects could lead not only to internal validity issues and non-reliable results but also to invalidate the experiment as *datasets randomization and dataset control strategies are a vital requirement of MSR experiments for ensuring their potential ability to deduce causality*.

## 4.- Be careful to claim causation and to generalize your results

All in all, our results denote that most MSR studies labeled as experiments overlook not only datasets randomization but also other relevant control strategies for MSR genuine experiments. In addition, the effect of these omissions on the level of internal validity and causality that can be reached from the studies has been also overlooked. We remark that the level of internal validity and causality that can be reached from an MSR Interventional Study (i.e., an MSR study fulfilling manipulation and datasets randomization) depends on its control nuances. ***The highest level of internal validity and causality can only be claimed from controlled experiments.***

Regarding generalization, although we found that 86% of MSR primary studies use multiple repositories to strength the external validity of their studies, most MSR primary studies selected their repositories based on convenience or availability (i.e., repositories selection was not random). Therefore, under such not random sampling selection conditions, we recommend MSR researchers to avoid generalizing their results outside the scope of the used repositories and the setting of the study [4],[36]. In other words, ***MSR researchers should be cautious about generalizing their results outside the scope of the used repositories unless the representativeness of the repositories can be justified.*** In addition, remember that nonrepresentative sampling should be followed by acknowledging that external validity is limited [4].

## 5.- Bear in mind the crucial methodological requirements of MSR study types and their implications on the appropriate use of the term *experiment*

To summarize most of this paper's recommendations, Table 8 provides a characterization of MSR study types based on their purpose, experimental design requirements and control alternatives. Regarding control, the effect of using prospective/retrospective repositories on the internal validity and causality than can be reached from each type of study is emphasized. In addition, an appropriate use the term *experiment* according to the characteristics of the MSR study type is suggested. This might serve as a ***quick an easy guidance for MSR researchers to apply most of the recommendations provided in this paper.***

Table 8 denotes, on the one hand, that those studies which purpose is to uncover potential associations and patterns are MSR Observational studies. In line with their purpose, observational studies do not cover the manipulation hallmark, so they should not be labeled as experiments as it would lead to a wrong use of the term. When designing this type of studies, researchers should keep in mind that MSR Observational studies have very flexible requirements regarding datasets randomization and control (if any), and they are usually retrospective. Although they are not able to detect causality (it is not their purpose), researchers should envisage a proper design that reinforces the internal validity and reliability of the study [61], [63]. The importance of observational studies has been widely recognized [21], [64], and they provide usually a solid basis to focus the design of an experiment as they help to identify trends and hypothesis to be tested [18], [64].

TABLE 8.
CHARACTERIZATION OF MSR STUDIES FROM A METHODOLOGICAL PERSPECTIVE

| MSR Purpose | MSR Study Type | Experimental Design Requirements to cover the Hallmarks of MSR Experiments | | | Internal validity | Causality | Appropriate use of the term **experiment** |
|---|---|---|---|---|---|---|---|
| | | *Dataset randomization* | Control | | | | |
| | | | *Dataset control Strategy* | *Dataset meas. & collection control* | | | |
| Uncovering associations and patterns | Observational | Could have | Could have | Usually Retrospective | Depends on the study design | NA | **Wrong use** |
| Comparing the behavior of different approaches under the same conditions | Interventional Study based on Comparison | ✓ | Dataset Blocking | Prospective | High | High | Controlled Experiment |
| | | | | Retrospective | Medium | Medium | Experiment with limited control |
| Training a ML algorithm to build an optimal resulting model | Interventional study based on Training ML algorithms | ✓ | Training/testing datasets + Dataset Repetition | Prospective | High | High | Controlled Experiment |
| | | | | Retrospective | Medium | Medium | Experiment with limited control |

On the other hand, studies aimed to compare the behavior of different approaches under the same conditions, as well as those aimed to train ML algorithms to calibrate an optimal resulting model, both fall into MSR Interventional studies category. Given their purposes, MSR Interventional studies cover the manipulation hallmark and can be considered as experiments. In addition to datasets randomization, MSR experiments require specific considerations regarding their experimental design: studies that aim to compare the behavior of diverse approaches require dataset blocking, while studies aimed to train ML algorithms require random selection of training/test datasets, and datasets repetition as default datasets control strategies. Furthermore, the level of control of MSR experiments could vary based on the use of prospective or retrospective repositories. The use of prospective repositories is the only way to guarantee the maximum level of control required by an *MSR controlled experiment*. The use of retrospective repositories leads to *MSR Experiments with limited control*. Such control nuances are relevant to properly interpret the results of MSR experiments.

The results and recommendations presented in this paper might help MSR researchers to better understand the peculiarities and methodological needs of MSR experiments and so they can pay special attention to strengthen those required needs that have been overlooked. It would improve the design, execution and reporting of MSR experiments.

## 6.- Be Aware of the Generic Nature of Current Experiment Guidelines used in MSR

MSR researchers use to inspire mainly on machine learning literature (e.g., [3], [36], [64]) and/or available SE guidelines for conducting experiments [28], [66]. In addition, best practices and bad smells have been shared in the MSR literature (e.g., [23], [39], [70]). However, although these recommendations and guidelines are quite useful for supporting the design, execution and reporting of MSR experiments under rigorous and systematic experimentation; none of these guidelines have been properly adapted to the specific characteristics and methodological needs of MSR research [72]. Therefore, these guidelines tend to overlook relevant aspects such as the ones discussed in this paper. To the best of our knowledge, there are no proper methodological guidelines yet to support MSR researchers (and reviewers) to design, execute and report their studies. So, we hope to raise the awareness of empirical SE researchers on the importance of such specific guidelines for fostering the proper use of SE empirical methods and a shared vocabulary in MSR (See recommendation 7). These specific guidelines are required not only for the diverse types of MSR experiments but also for MSR Observational studies, as there are in other disciplines such as medicine [43], [63] or bioinformatics [15].

### 7.- Recognition of the diversity of empirical studies in SE.

Finally, we would like to raise this final recommendation for the SE community. The lack of clear methodological guidelines for MSR studies leads to confusion not only on the type of evidence that can be obtained from them but also on the use of SE empirical terminology. In this paper we focused on the assessment of MSR experiments, but we observed that the terminology confusion seems general as some authors of MSR primary studies, in addition to referring to their MSR studies as "experiments", they also call them "case studies" (for instance, [MSR2016-1], [EMSE2017-64], [EMSE2017-29]). It seems that some researchers consider that the use of different repositories might be similar to "cases"; however, it does not really fit the current definition of a "case study" in SE: *"case study is an empirical method aimed at investigating contemporary phenomena in their context [...] using multiple sources of evidence. It refers to an inquiry where the boundary between the phenomenon and its context may be unclear and lack of experimental control"* [51].

We hope that our results foster the SE community to recognize the need of further efforts to understand the peculiarities of MSR studies and to accommodate them into the current SE empirical methods. We think that it is a required step for maturing the conception of SE empirical studies in general and MSR studies in particular. Similar efforts have been done in medical research to reconcile the great variety of empirical studies they perform [50]. We think that the recognition and development of epidemiology as a fundamental part of medical research seems a suitable analogy to MSR research in SE that is worth to explore. It is because epidemiology is also a data-driven discipline that relies on a systematic and unbiased approach to the collection, analysis, and interpretation of data [40], [47], [63].

## 8. THREATS TO VALIDITY

Our results are based on the assessment of MSR literature from a small-scale systematic mapping study using well-known guidelines and recommendations [7], [33]. Moreover, the authors have proved experience on performing systematic literature studies.

As all studies, our literature study has some relevant limitations. First, the consideration of only three venues (MSRConf, EMSE, ESEM) and four years (2015-2018) could increase the risk of not covering all possible nuances of the concept of experiment as used by the MSR community. This was done because from a practical point of view, the type of investigation we need to conduct on every primary study avoids the analysis of a very large set of papers, therefore we considered appropriate to select a representative sample of venues that allow us to analyze in deep each primary study. To mitigate such risk, the selected venues are highly representative of the main publication channels for empirical SE [5] and MSR research [23]. The selected years provide recent information of the MSR literature. So, we do think that the results we get are representative enough to fulfill our goal of having a deep understanding on the use of the term experiment in MSR research.

Another threat to this review is the possible inaccuracy and subjectivity in data extraction and classification of the studies. On the one hand, the data was extracted mainly by one person (the first author). However, the second and fourth author randomly choose papers to confirm the extracted/classified data. Any differences between the reviewers were solved through discussions until a consensus was reached. We did not experience critical differences. On the other hand, in general, our research team has a well-known expertise and background on empirical SE and MSR research and we had several long discussion meetings to devise the assessment criteria devised in the study. This helps to mitigate the risks of misunderstandings on devising and using the criteria.

## 9. CONCLUSIONS

The work reported here aims to raise the awareness of the special characteristics of MSR experiments that make them different from traditional SE experiments. Our results evidenced that the use of the term experiment in MSR is problematic: 19% of the assessed papers use the term experiment in a wrong way as they are not experiments at all but also observational studies. From the remaining 81% of the papers, most of them overlook not only datasets randomization but also relevant dataset control strategies as well as the effect of these omissions on the limitations of the studies. We provided recommendations and insights to support the improvement of MSR experiments. Table 8 summarizes most of our recommendations and can be used as a preliminary guide for understanding the experimental design requirements of MSR studies. In addition, it would help to raise the awareness of the experimental design implications on the internal validity and causality that can be reached from each type of MSR study, as well as appropriate uses of the term experiment.

Our results might help to:

a) MSR researchers to better understand the peculiarities and methodological needs of MSR experiments; and to improve the design, execution and reporting of MSR experiments by avoiding overlooking critical aspects that we have highlighted.

b) the SE community to raise the awareness on the importance of supporting MSR research with proper methodological guidelines that help to smoothly reconcile use of SE empirical methods and a shared vocabulary.

c) Reviewers and readers of MSR papers can use the recommendations and characterization of MSR studies provided in Table 8 for supporting their interpretation of the results of MSR studies and/or evaluate their methodological aspects.

## ACKNOWLEDGEMENT

We are extremely grateful to Dr. Xin Xia for their valuable and constructive feedback.
This work has been partially supported by the Spanish project: MCI PID2020-117191RB-I00.

## PRIMARY STUDIES

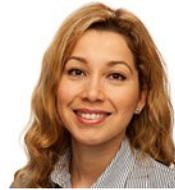

**Claudia Ayala** is an Associate Professor of software engineering at Universitat Politècnica de Catalunya (UPC-BarcelonaTech). She received her PhD degree in Software from UPC in 2008. She was a Post-Doctoral Fellow of the European Research Consortium for Informatics and Mathematics (ER-CIM) at the Norwegian University of Science and Technology (NTNU), Norway 2008-2009.  She is regular reviewer of highly ranked journals such as IEEE Transactions on Software Engineering, Empirical Software Engineering Journal, and Information and Software Technology. Dr. Ayala has actively participated in the SE community as Project Manager of ICSE 2021, Program Co-chair for the Ibero-American Conference on Software Engineering (CIbSE) and several other events such as Posters and Demos Track-RCIS 2017; Short papers and posters track-EASE 2015; ESELAW-CIbSE 2019; Proceedings (Co)Chair- PROFES 2019, CAiSE'12; Student Volunteer Chair-RE 2008. Her current research interests include empirical software engineering, requirements engineering, software architecture and quality, and open source software adoption. More information at: http://www.essi.upc.edu/~cayala/

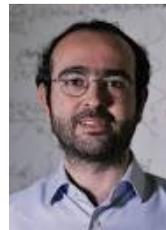

**Burak Turhan** PhD (Boğaziçi University), is a Professor of Software Engineering at the University of Oulu, Finland, and an Adjunct Professor (Research) in the Faculty of IT at Monash University, Australia. His research focuses on empirical software engineering, software analytics, quality assurance and testing, human factors, and (agile) development processes. He is a Senior Associate Editor of the Journal of Systems and Software, an Associate Editor of ACM Transactions on Software Engineering and Methodology and Automated Software Engineering, an Editorial Board Member of Empirical Software Engineering, Information and Software Technology, and Software Quality Journal, and a Senior Member of ACM and IEEE. For more information, please visit:https://turhanb.net.

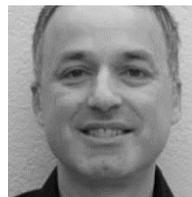

**Xavier Franch** is a full professor in Software Engineering at the Universitat Politècnica de Catalunya (UPC-BarcelonaTech). He received his PhD degree in Informatics from UPC in 1996. His research interest embraces many fields in software engineering, including requirements engineering, empirical software engineering, open source software, and agile software development. Prof. Franch is a member of the IST, REJ, IJCIS, and Computing editorial boards, Journal First chair of JSS, and Deputy Editor of IET Software. He served as a PC chair at RE'16, ICSOC'14, CAiSE'12, and REFSQ'11, among others, and as GC for RE'08 and PROFES'19. More information at https://www.essi.upc.edu/~franch.

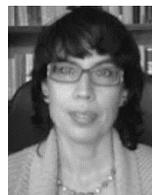

**Natalia Juristo** has been full professor of software engineering with the School of Computer Engineering, Technical University of Madrid (UPM) since 1997. She was awarded a FiDiPro (Finland Distinguished Professor Program) professorship with the University of Oulu and she was also awarded an honorary doctorate by Blekinge Institute of Technology in Sweden. She was the director of the MSc in software engineering and coordinator of the Erasmus Mundus European Master on SE (with the participation of UPM, University of Bolzano, University of Kaiserslautern and Blekinge Institute of Technology) from 2006 to 2012. Her main research interests include experimental software engineering, requirements and testing. She co-authored the book Basics of Software Engineering Experimentation (Kluwer) and is a member of the editorial boards: IEEE Transactions on Software Engineering, Empirical SE, and Software: Testing, Verification and Reliability. She has served on several congress program committees (ICSE, RE, REFSQ, ESEM, ISESE, etc.), and has been congress program chair (EASE13, ISESE04 and SEKE97), as well as general chair (ICSE 2021, ESEM07, SNPD02 and SEKE01). More on http://grise.upm.es/miembros/natalia/